\documentclass[11pt]{article}
\usepackage{epsfig} 
\newcommand{\sect}[1]{ \section{#1} \setcounter{equation}{0} }

\newcommand{\Dslash}{D \! \! \! \! /}

\newcommand{\half}{\mbox{\small{$\frac{1}{2}$}}} 
\newcommand{\Nc}{N_{\!c}} 
\newcommand{\Nf}{N_{\!f}} 
\newcommand{\MSbar}{\overline{\mbox{MS}}}

\begin{document}
\title{Two loop effective potential for $\left\langle A^2_\mu \right\rangle$ in
the Landau gauge in quantum chromodynamics} 
\author{R.E. Browne \& J.A. Gracey, \\ Theoretical Physics Division, \\ 
Department of Mathematical Sciences, \\ University of Liverpool, \\ P.O. Box 
147, \\ Liverpool, \\ L69 3BX, \\ United Kingdom.} 
\date{} 
\maketitle 
\vspace{5cm} 
\noindent 
{\bf Abstract.} We construct the effective potential for the dimension two
composite operator $\half A^{a \, 2}_\mu$ in QCD with massless quarks in the 
Landau gauge for an arbitrary colour group at two loops. For $SU(3)$ we show 
that an estimate for the effective gluon mass decreases as $\Nf$ increases.  

\vspace{-15.5cm}
\hspace{13.5cm} 
{\bf LTH 581} 

\newpage

\sect{Introduction.} 
Quantum chromodynamics, (QCD), is widely accepted as the quantum field theory
of the nucleon partons known as quarks and gluons. At high energy, perturbation
theory provides an excellent description of the phenomenology of hadron physics
from, say, the point of view of deep inelastic scattering. Although the physics
of QCD at low energies is not as well understood, it is generally accepted 
that quarks are confined and the perturbative vacuum, which is used
for higher energy QCD calculations, is unstable, \cite{1,2,3,4}. To probe the
infrared r\'{e}gime from the field theoretic point of view several formalisms
have been developed. One in particular makes use of the operator product
expansion and sum rules where the effect of dimension four operators such as
$(G^a_{\mu\nu})^2$ are incorporated, \cite{5}. Such operators have non-zero 
vacuum expectation values and can be regarded as probing the structure of the 
true vacuum. Such methods have been successful in revealing insights into the 
low energy structure of Yang-Mills theory and QCD. However, in this general 
context there has been a large amount of activity recently into the effects 
that dimension two operators have on this scenario,
\cite{6,7,8,9,10,11,12,13,14,15,16,17,18,19}. For instance, the operator
$\half A^{a \, 2}_\mu$ and the related BRST invariant operator  
$\half A^{a \, 2}_\mu$~$+$~$\alpha \bar{c}^a c^a$ have received particular
attention being motivated by early work on the Curci-Ferrari model, \cite{20}, 
where $A^a_\mu$ is the gluon field and $c^a$ and $\bar{c}^a$ are the ghost and 
anti-ghost fields of the covariant gauge fixing which is parametrized by 
$\alpha$. The BRST algebraic properties of these operators has been extensively
studied, \cite{21,22,23,24,25,26,27,28}, particularly in relation to unitarity. 
Unlike $(G^a_{\mu\nu})^2$ such operators are clearly not gauge invariant and 
therefore they can only arise in gauge variant quantities such as the strong 
coupling constant. (See, for example, \cite{29,30}.) However, one can construct
a dimension two gauge invariant operator which is non-local, \cite{11}, but is 
believed to reduce to $\half A^{a \, 2}_\mu$ in the Landau gauge where it is 
clearly local. Therefore, various groups have examined the effect such an 
operator has in Yang-Mills theories both in the case of ordinary linear
covariant gauge fixing and in a related non-linear covariant gauge fixing. The
latter is known as the Curci-Ferrari gauge having its origin in the 
Curci-Ferrari model which was an early attempt to construct a field theory
where the gluon and ghosts were classically massive, \cite{20}. This gauge 
differs from the usual covariant gauge fixing due to the presence of a 
renormalizable four-ghost interaction and a different ghost-gluon interaction. 
Its ultra-violet behaviour is equivalent to that of the usual linear Landau 
gauge when the covariant gauge parameter is nullified. 

One important study of the effects of the $\half A^{a \, 2}_\mu$ operator has  
been provided in \cite{11,31}. There the effective potential of this operator 
has been constructed at two loops in Yang-Mills theory. This is not a 
straightforward exercise as one has to apply the local composite operator, 
(LCO), formalism which correctly accounts for a composite operator being 
included in the path integral formalism, \cite{11,31,32,33}. Early work with 
this technique centred on studying mass generation in the two dimensional 
Gross-Neveu model where estimates for the mass gap were obtained which were in 
good agreement with the known exact mass gap of the model, \cite{32,33}. 
Consequently the LCO technique provides an effective action which corresponds 
to the Yang-Mills action as well as a new contribution which involves an extra 
scalar field whose elimination by its equation of motion is related to the 
dimension two composite operator. In this respect it is similar to the 
Gross-Neveu model though in that case the analogous scalar field arises 
{\em naturally} in the Lagrangian. With the new action derived with the 
LCO technique the effective potential of the dimension two operator has been 
constructed at two loops in Yang-Mills theory, \cite{11}. However, as the 
nucleon world possesses quarks there is clearly a need to repeat the analysis 
for full QCD. This is the main purpose of this article where we will extend the
effective potential of \cite{11} to include massless quarks. Whilst it may 
appear that our calculation follows \cite{11}, it should be noted that 
\cite{11} considered only the $SU(\Nc)$ colour groups and, moreover, required 
the evaluation of three loop massive vacuum bubble graphs using the tensor 
correction method developed in \cite{31}. By contrast, we will consider
arbitrary colour groups which allows one to study, in principle, the effect
in grand unified theories. More importantly, though, we will bypass the need
to compute three loop massive vacuum bubble diagrams. These were required
to determine the explicit expression for a particular parameter, $\zeta(g)$,
as a function of the coupling constant, $g$. This quantity, which plays a
role similar to the coupling constant for the new scalar field of the new
action, obeys a particular renormalization group equation and its explicit
value, as we will demonstrate, only requires the evaluation of three loop
{\em massless} Feynman diagrams. The importance of this observation is not to 
be underestimated since from a calculational point of view it reduces the work
significantly and will be important in other similar applications of the LCO
formalism for which the symbolic manipulation programmes developed here can be
applied. One final motivation for this article is to ascertain the effect the 
quarks have on an effective gluon mass. In the construction of the two loop 
effective potential it transpires that for both Yang-Mills and QCD the naive 
perturbative vacuum is not stable, \cite{11}, but there is a stable vacuum 
where the gluon field develops a mass through the non-zero vacuum expectation 
value of the new scalar field arising from the LCO formalism. Clearly an 
estimate of such a mass is important when quarks are present. While our 
calculations are still effectively perturbation theory probing towards the true
non-perturbative vacuum, the full vacuum expectation value of the scalar field 
will be comprised of two parts. One is the perturbative piece estimated from 
the effective potential but the other component will arise from purely 
non-perturbative phenomena such as instantons, \cite{11}. In this context
there has been various studies of dimension two operators and condensates on 
the lattice giving estimates of similar quantities, \cite{29,30}. Hence, it 
will be useful to provide information with quarks present ahead of lattice 
simulations of full QCD where it would be hoped the dimension two condensates 
could be measured at an energy scale compatible with the energy scale we will 
determine our effective gluon mass estimates.

The paper is organised as follows. In section two we will review the 
construction of the LCO formalism for full QCD and determine several quantities
which are central to the construction of the two loop effective potential in
the QCD case. The full two loop potential is discussed in section three whilst
we provide a detailed analysis of the colour groups $SU(2)$ and $SU(3)$ in
section four. Concluding remarks are given in section five.  

\sect{Formalism.}
In this section we review the basic formalism for constructing the effective
potential of a composite operator using the LCO formalism, \cite{11}. Our 
notation parallels that of \cite{11} and we will focus on the main ingredients
as well as where new features emerge. The starting point is the construction
of an effective action which includes the composite operator we are interested
in, $\half A^{a \, 2}_\mu$, in the Landau gauge, and is based on the usual
QCD action. In particular the QCD Lagrangian is  
\begin{equation} 
L ~=~ -~ \frac{1}{4} G_{\mu\nu}^a G^{a \, \mu\nu} ~-~ \frac{1}{2\alpha} 
(\partial^\mu A^a_\mu)^2 ~-~ \bar{c}^a \partial^\mu D_\mu c^a ~+~ 
i \bar{\psi}^{iI} \Dslash \psi^{iI} 
\label{lag}
\end{equation} 
where $G^a_{\mu\nu}$ $=$ $\partial_\mu A^a_\nu$ $-$ $\partial_\nu A^a_\nu$ $-$
$g f^{abc} A^b_\mu A^c_\nu$, $f^{abc}$ are the colour group structure 
constants, $\psi^{iI}$ is the quark field and $1$~$\leq$~$a$~$\leq$~$N_A$,
$1$~$\leq$~$I$~$\leq$~$N_F$ and $1$~$\leq$~$i$~$\leq$~$\Nf$ with $N_F$ and 
$N_A$ the dimensions of the fundamental and adjoint representations 
respectively and $\Nf$ is the number of quark flavours. The covariant 
derivatives are defined by  
\begin{equation} 
D_\mu c^a ~=~ \partial_\mu c^a ~-~ g f^{abc} A^b_\mu c^c ~~,~~ 
D_\mu \psi^{iI} ~=~ \partial_\mu \psi^{iI} ~+~ i g T^a A^a_\mu \psi^{iI} ~.  
\end{equation} 
Although most of the recent activity in this area has dwelt on Yang-Mills in
the Curci-Ferrari gauge since it has properties in common with the maximal
abelian gauge, we have chosen to fix the gauge with the usual linear covariant 
gauge fixing. This is because the potential in either this gauge or the 
Curci-Ferrari gauge is the same when $\alpha$ $=$ $0$. For reasons which will 
become apparent later the quarks in (\ref{lag}) are massless. Moreover, we can 
regard (\ref{lag}) as the bare Lagrangian and introduce renormalized quantities
via the usual definitions such as 
$A^{a \, \mu}_{\mbox{\footnotesize{o}}}$~$=$~$\sqrt{Z_A} \, A^{a \, \mu}$ and 
$g_{\mbox{\footnotesize{o}}}$~$=$~$Z_g g$ where the subscript, 
${}_{\mbox{\footnotesize{o}}}$, denotes a bare quantity.  

To include the composite operator, $\half A^{a \, 2}_\mu$, in the path integral 
it would initially seem that one should include the term 
$\half J A^{a \, 2}_\mu$ in the Lagrangian where $J$ is a source term and
derive the usual path integral generating functional, $W[J]$. Ordinarily in
such an approach one couples the fields themselves to each source to obtain a
generating functional which is used to determine the Green's functions.
However, in coupling to an object quadratic in the fields an immediate problem
arises in that the new action ceases to be multiplicatively renormalizable,
\cite{11}. This is due to the generation of divergences proportional to $J^2$
which are related to vacuum energy divergences, \cite{11}. Divergences 
involving one power of $J$ are absorbed into the renormalization constant for 
the renormalization of the composite operator, $\half A^{a \, 2}_\mu$, itself,
\cite{31,34}. Similar features are also present in the Gross-Neveu model,
\cite{32,35}. To circumvent this lack of multiplicative renormalizability one 
must add a term quadratic in $J$ with an appropriate counterterm in order to 
have a sensible generating functional, $W[J]$. Therefore, the generating 
functional which is the starting point of the LCO formalism is, \cite{11},  
\begin{equation}
e^{-W[J]} ~=~ \int {\cal D} A^\mu_{\mbox{\footnotesize{o}}} 
{\cal D} \psi_{\mbox{\footnotesize{o}}} {\cal D} 
\bar{\psi}_{\mbox{\footnotesize{o}}} {\cal D} c_{\mbox{\footnotesize{o}}} 
{\cal D} \bar{c}_{\mbox{\footnotesize{o}}} \,
\exp \left[ \int d^d x \left( L_{\mbox{\footnotesize{o}}} ~-~ \frac{1}{2} 
J_{\mbox{\footnotesize{o}}} A_{{\mbox{\footnotesize{o}}} \, \mu}^{a \, 2} ~+~ 
\frac{1}{2} \zeta_{\mbox{\footnotesize{o}}} J_{\mbox{\footnotesize{o}}}^2 
\right) \right]
\label{bareW}
\end{equation}  
where all quantities are bare and the new quantity 
$\zeta_{\mbox{\footnotesize{o}}}$ has been introduced to ensure one will have a 
homogeneous renormalization group equation for
$W[J]$, \cite{11}, whose conventions we follow throughout. 

With the extra terms in the action additional renormalization constants are
required over and above those needed to render the normal QCD Lagrangian, $L$,
finite, \cite{11}. However, it is important to realise that the additional 
terms do not affect the explicit values of the ordinary QCD renormalization 
group functions. This can be understood from the fact that the new cubic 
interaction can be regarded as a mass term for the gluon which, from the point 
of view of renormalization, will regularize infrared infinities and leave the 
ultraviolet structure unchanged. Therefore, we need to introduce two new 
renormalization constants which to be consistent with \cite{11} we denote by 
$Z_m$ and $\delta \zeta$ where the latter is strictly a counterterm. Therefore,
using 
\begin{equation} 
J_{\mbox{\footnotesize{o}}} ~=~ \frac{Z_m}{Z_A} J ~~~,~~~ 
\zeta_{\mbox{\footnotesize{o}}} J_{\mbox{\footnotesize{o}}}^2 ~=~ \left( 
\zeta ~+~ \delta \zeta \right) J^2 
\end{equation} 
(\ref{bareW}) becomes 
\begin{equation}
e^{-W[J]} ~=~ \int {\cal D} A_\mu {\cal D} \psi {\cal D} \bar{\psi} {\cal D} c 
{\cal D} \bar{c} \, \exp \left[ \int d^d x \left( L ~-~ \frac{1}{2} Z_m J 
A_\mu^{a \, 2} ~+~ \frac{1}{2} ( \zeta + \delta \zeta ) J^2 \right) \right] ~. 
\label{renW}
\end{equation}  
For the construction of the effective potential the explicit values of each
renormalization constant will be required. It turns out that these are
straightforward to compute. For instance, $Z_m$ has already been 
determined in \cite{26,36,37,38,39} at one and two loops and to three loops in
\cite{31,34}. The method to achieve this is twofold. One can either use a
Lagrangian with a gluon of squared mass $J$ and renormalize with a massive
gluon propagator in the Landau gauge or one can regard the operator  
$\half J A^{a \, 2}_\mu$ as a composite operator and renormalize it as an
operator insertion in a gluon two-point function using {\em massless}
propagators. Indeed it was the latter approach which was used to deduce the 
three loop operator anomalous dimension \cite{34}. However, a key property that 
emerged from that work, which has subsequently been proved to all orders in 
\cite{17}, is that the renormalization of $\half A^{a \, 2}_\mu$ is not 
independent. In the Landau gauge the anomalous dimension is given by the sum of
the gluon and ghost wave function anomalous dimensions. As this will play an 
important role in the construction of the effective potential we note that in 
the Landau gauge the explicit value in the $\MSbar$ scheme is 
\begin{eqnarray} 
\left. \gamma_m(g) \right|_{\alpha \, = \, 0} &=& \left[ \frac{35}{6} C_A 
- \frac{8}{3} T_F \Nf \right] \frac{g^2}{16\pi^2} ~+~ \left[ \frac{449}{24} 
C_A^2 - 8 C_F T_F \Nf - \frac{35}{3} C_A T_F \Nf \right] 
\frac{g^4}{(16\pi^2)^2} \nonumber \\  
&& +~ \left[ \left( \frac{75607}{864} - \frac{9}{16} \zeta(3) \right) C_A^3
+ \frac{88}{9} T_F^2 \Nf^2 C_F + \frac{386}{27} T_F^2 \Nf^2 C_A 
+ 4 T_F \Nf C_F^2 \right. \nonumber \\
&& \left. ~~~~~+~ \left( 18 \zeta(3) - \frac{5563}{54} \right) T_F \Nf C_A^2
- \left( 24 \zeta(3) + \frac{415}{18} \right) T_F \Nf C_F C_A \right] 
\frac{g^6}{(16\pi^2)^3} \nonumber \\
&& +~ O(g^8) 
\label{gammam}
\end{eqnarray} 
where the Casimirs are defined by $T^a T^a$ $=$ $C_F I$, $f^{acd} f^{bcd}$ $=$
$C_A \delta^{ab}$ and $\mbox{Tr}\left( T^a T^b \right)$ $=$ $T_F \delta^{ab}$.

Therefore all that remains is to evaluate the counterterm $\delta \zeta$. In
\cite{11} the source, $J$, was considered to be constant giving the gluon a
mass. The counterterm was then deduced by renormalizing three loop massive 
vacuum bubble diagrams generated from the interactions of $L$. This required 
the development of the tensor correction method, \cite{31}, to efficiently 
handle the tensor reduction and algebra associated with massive Feynman 
diagrams. Then the divergences of the vacuum diagrams, which were proportional 
to $J^2$, were absorbed into $\delta \zeta$. In our approach we have considered 
the source, $J$, to be a field which interacts with the gluon leaving no 
candidate for a mass term for the gluon. It therefore remains massless. By 
viewing the source like an interaction rather than a mass, we have avoided the 
need to evaluate massive vacuum bubbles. Instead we proceed by considering the 
vacuum Green's function  
\begin{equation} 
\langle 0 | {\cal O}(x) \, {\cal O}(y) | 0 \rangle
\label{OOgf} 
\end{equation}
using (\ref{lag}) where ${\cal O}(x)$ $=$ $\half J A^{a \, 2}_\mu$. It is the
divergences of (\ref{OOgf}) which must be cancelled by $\delta \zeta$. We 
carried out the three loop renormalization by using the {\sc Mincer} algorithm,
\cite{40}, as implemented in {\sc Form}, \cite{41,42}, with the relevant 
Feynman diagrams generated by {\sc Qgraf}, \cite{43}. The converter files 
mapping the {\sc Qgraf} output into input {\sc Form} notation have already been
used in \cite{34}. However, the {\sc Mincer} algorithm evaluates the 
divergences of two-point functions whereas our vacuum Green's function has no
external legs. It was therefore necessary to cut each diagram through the one
internal $J$ propagator and consider the resulting $J$ two-point function.
This is valid because the propagator of this source field is a constant and is
thus common to each graph. It is important to note that this is not equivalent
to considering the two-point source Green's function for the complete
Lagrangian, (\ref{renW}), as this would contain extra diagrams with internal
$J$ propagators which do not contribute to (\ref{OOgf}). From a calculational 
perspective it would seem that renormalizing (\ref{OOgf}) in this way is more 
efficient since one needs only to use {\em massless} propagators. Indeed this
alternative could probably be applied to similar computations of the effective 
potentials of other dimension two composite operators. One benefit, for 
instance, of using the {\sc Mincer} algorithm, \cite{40,41}, is that the 
tedious tensor reductions are automatically implemented. To check that this 
interpretation is consistent with the three loop massive vacuum bubble 
calculation of \cite{11,31} we have renormalized (\ref{OOgf}) for arbitrary 
$\Nf$ and found in dimensional regularization with $d$~$=$~$4$~$-$~$2\epsilon$ 
using $\MSbar$, that 
\begin{eqnarray}
\delta \zeta &=& N_A \left[ ~-~ \frac{3}{2\epsilon} ~+~ \left( \left( 
\frac{35}{8} C_A - 2 T_F \Nf \right) \frac{1}{\epsilon^2} + \left( \frac{8}{3}
T_F \Nf - \frac{139}{12} C_A \right) \frac{1}{\epsilon} \right) 
\frac{g^2}{16\pi^2} \right. \nonumber \\
&& \left. ~~~~~~~+~ \left( \left( \frac{73}{6} T_F \Nf C_A - \frac{8}{3} T_F^2 
\Nf^2 - \frac{665}{48} C_A^2 \right) \frac{1}{\epsilon^3} \right. \right.
\nonumber \\
&& \left. \left. ~~~~~~~~~~~~~+~ \left( \frac{32}{9} T_F^2 \Nf^2 
- 4 T_F \Nf C_F - \frac{535}{18} T_F \Nf C_A + \frac{6629}{144} C_A^2 \right) 
\frac{1}{\epsilon^2} \right. \right. \nonumber \\
&& \left. \left. ~~~~~~~~~~~~~+~ \left( \frac{40}{27} T_F^2 \Nf^2 
+ \left( \frac{115}{3} - 32 \zeta(3) \right) T_F \Nf C_F 
+ \left( \frac{4381}{216} + 32 \zeta(3) \right) T_F \Nf C_A \right. \right.
\right.  \nonumber \\
&& \left. \left. \left. ~~~~~~~~~~~~~-~ \left( \frac{71551}{864} 
+ \frac{231}{32} \zeta(3) \right) C_A^2 \right) 
\frac{1}{\epsilon} \right) \frac{g^4}{(16\pi^2)^2} \right] ~+~ O(g^6) ~. 
\label{zetaren}
\end{eqnarray} 
When $\Nf$ $=$ $0$ (\ref{zetaren}) corresponds {\em exactly} with the result of
\cite{11}\footnote{In \cite{11} the convention $d$~$=$~$4$~$-$~$\epsilon$ was 
used.}. 

With (\ref{zetaren}) we need to construct the associated renormalization group
function denoted by $\delta(g)$ and defined by 
\begin{equation}
\mu \frac{\partial \zeta}{\partial \mu} ~=~ 2 \gamma_m(g) \zeta ~+~ 
\delta(g) ~.  
\end{equation}
The renormalization group equation which $W[J]$ satisfies is, \cite{11},  
\begin{equation}
\left[ \mu \frac{\partial ~}{\partial \mu} ~+~ \beta(g) \frac{\partial ~}
{\partial g^2} ~-~ \gamma_m(g) \int_x J \frac{\delta ~}{\delta J} ~+~
\left[ \delta(g) + 2 \zeta \gamma_m(g) \right] \frac{\partial ~}{\partial 
\zeta} \right] W[J] ~=~ 0 ~.
\end{equation} 
Thus from 
\begin{equation}
\delta(g) ~=~ \left[ 2 \epsilon ~+~ 2 \gamma_m(g) ~-~ 
\beta(g) \frac{\partial~}{\partial g^2} \right] \delta \zeta
\end{equation} 
we find 
\begin{eqnarray}
\delta(g) &=& \frac{N_A}{16\pi^2} \left[ ~-~ 3 ~+~ \left( \frac{32}{3} T_F \Nf
- \frac{139}{3} C_A \right) \frac{g^2}{16\pi^2} \right. \nonumber \\
&& \left. ~~~~~~~~~+~ \left( \left( \frac{4381}{36} + 192 \zeta(3) \right) 
T_F \Nf C_A + \left( 230 - 192 \zeta(3) \right) T_F \Nf C_F 
+ \frac{80}{9} T_F^2 \Nf^2 \right. \right. \nonumber \\
&& \left. \left. ~~~~~~~~~-~ \left( \frac{71551}{144} + \frac{693}{16} \zeta(3)
\right) C_A^2 \right) \frac{g^4}{(16\pi^2)^2} \right] ~+~ O(g^6) ~. 
\end{eqnarray} 
The next stage, \cite{11}, is to choose $\zeta(g)$ to be the solution of the 
differential equation 
\begin{equation}
\beta(g) \frac{d~}{dg} \zeta(g) ~=~ 2 \gamma_m(g) \zeta(g) ~+~ \delta(g)
\label{zetarge}
\end{equation} 
so that the running of the coupling coupling constant means that $\zeta(g)$ 
will also run according to its renormalization group equation, (\ref{zetarge}),
and which ensures that one can construct a consistent effective action which
includes $\half J A^{a \, 2}_\mu$ as a composite operator. Given the nature
of this differential equation its solution is singular at the origin and has
the expansion, in powers of the coupling constant,  
\begin{equation}
\zeta(g) ~=~ \sum_{n \, = \, - \, 1}^\infty c_n g^{2n} ~.  
\end{equation} 
With the explicit expressions we have quoted we can deduce that  
\begin{eqnarray}
\frac{1}{g^2\zeta(g)} &=& \left[ \frac{( 13 C_A - 8 T_F \Nf )}{9N_A} 
\right. \nonumber \\
&& \left. +~ \left( 2685464 C_A^3 T_F \Nf - 1391845 C_A^4 
- 213408 C_A^2 C_F T_F \Nf - 1901760 C_A^2 T_F^2 \Nf^2 
\right. \right. \nonumber \\
&& \left. \left. ~~~~~ 
+~ 221184 C_A C_F T_F^2 \Nf^2 + 584192 C_A \Nf^3 T_F^3 
- 55296 C_F T_F^3 \Nf^3 
\right. \right. \nonumber \\
&& \left. \left. ~~~~~ 
-~ 65536 T_F^4 \Nf^4 \right) \frac{g^2}{5184 \pi^2 N_A (35 C_A-16 T_F \Nf) 
(19 C_A-8 T_F \Nf)} 
\right. \nonumber \\
&& \left. 
+~ \left( \left( 62228252520 C_A^6 \Nf T_F - 8324745975 C_A^7 
- 42525100800 C_A^5 C_F \Nf T_F 
\right. \right. \right. \nonumber \\
&& \left. \left. \left. ~~~~~ 
-~ 123805256256 C_A^5 \Nf^2 T_F^2 + 105262940160 C_A^4 C_F \Nf^2 T_F^2
\right. \right. \right. \nonumber \\
&& \left. \left. \left. ~~~~~ 
+~ 112398515712 C_A^4 \Nf^3 T_F^3 
- 103719518208 C_A^3 C_F \Nf^3 T_F^3 
\right. \right. \right. \nonumber \\
&& \left. \left. \left. ~~~~~ 
-~ 52888043520 C_A^3 \Nf^4 T_F^4 + 50866421760 C_A^2 C_F \Nf^4 T_F^4
\right. \right. \right. \nonumber \\
&& \left. \left. \left. ~~~~~ 
+~ 12606898176 C_A^2 \Nf^5 T_F^5 - 12419334144 C_A C_F \Nf^5 T_F^5
\right. \right. \right. \nonumber \\
&& \left. \left. \left. ~~~~~ 
-~ 1207959552 C_A \Nf^6 T_F^6 + 1207959552 C_F \Nf^6 T_F^6 \right) \zeta(3)
- 13223737800 C_A^7 
\right. \right. \nonumber \\
&& \left. \left. ~~~~~ 
+~ 5886241060 C_A^6 \Nf T_F + 52585806000 C_A^5 C_F \Nf T_F 
+ 41351916768 C_A^5 \Nf^2 T_F^2 
\right. \right. \nonumber \\
&& \left. \left. ~~~~~ 
+~ 522849600 C_A^4 C_F^2 \Nf T_F - 130596636288 C_A^4 C_F \Nf^2 T_F^2 
\right. \right. \nonumber \\
&& \left. \left. ~~~~~ 
-~ 67857620736 C_A^4 \Nf^3 T_F^3 - 1286267904 C_A^3 C_F^2 \Nf^2 T_F^2 
\right. \right. \nonumber \\
&& \left. \left. ~~~~~ 
+~ 128750638080 C_A^3 C_F \Nf^3 T_F^3 + 46700324864 C_A^3 \Nf^4 T_F^4 
\right. \right. \nonumber \\
&& \left. \left. ~~~~~ 
+~ 1180127232 C_A^2 C_F^2 \Nf^3 T_F^3 - 63001780224 C_A^2 C_F \Nf^4 T_F^4 
\right. \right. \nonumber \\
&& \left. \left. ~~~~~ 
-~ 16782753792 C_A^2 \Nf^5 T_F^5 - 475987968 C_A C_F^2 \Nf^4 T_F^4 
\right. \right. \nonumber \\
&& \left. \left. ~~~~~ 
+~ 15308685312 C_A C_F \Nf^5 T_F^5 + 3106406400 C_A \Nf^6 T_F^6 
\right. \right. \nonumber \\
&& \left. \left. ~~~~~ 
+~ 70778880 C_F^2 \Nf^5 T_F^5 - 1478492160 C_F \Nf^6 T_F^6 
\right. \right. \nonumber \\
&& \left. \left. ~~~~~ 
-~ 234881024 \Nf^7 T_F^7 \right) 
\frac{g^4}{995328 \pi^4 N_A (35 C_A-16 T_F \Nf)^2 (19 C_A-8 T_F \Nf)^2} 
\right] \nonumber \\
&& +~ O(g^6)
\label{zetaval} 
\end{eqnarray}
where the $\Nf$ $=$ $0$ case agrees {\em exactly} with the corrected result of
the erratum of \cite{11}. 

Finally, for practical purposes it is more appropriate to transform the action,
$W[J]$, to a situation where it depends linearly on the source, $J$, which
will allow one to apply the usual path integral machinery to construct the
effective potential itself. In \cite{11} this was achieved by the 
Hubbard-Stratonovich transformation where one includes the redundant term 
\begin{equation}
1 ~=~ \int {\cal D} \sigma \, \exp \left( -~ \left[ a_1 \sigma ~+~ 
a_2 A^{a \, 2}_\mu ~+~ a_3 J \right]^2 \right) 
\end{equation} 
in the path integral where $\sigma$ is a scalar auxiliary field. The arbitrary 
coefficients, $\{a_i\}$, are chosen appropriately to cancel off the $\half J^2$
and $\half J A^{a \, 2}_\mu$ terms to leave a term proportional to $\sigma J$ 
as the only source dependence. These coefficients will depend on $g$, 
$\zeta(g)$, $Z_m$ and $Z_\zeta$. This results in a new Lagrangian, involving 
the field $\sigma$, which includes the original QCD Lagrangian with its gauge
fixing and a new set of interactions  
\begin{eqnarray} 
L^\sigma &=& -~ \frac{1}{4} G_{\mu\nu}^a G^{a \, \mu\nu} ~-~ \frac{1}{2\alpha} 
(\partial^\mu A^a_\mu)^2 ~-~ \bar{c}^a \partial^\mu D_\mu c^a ~+~ 
i \bar{\psi}^{iI} \Dslash \psi^{iI} \nonumber \\ 
&& -~ \frac{\sigma^2}{2g^2 \zeta(g) Z_\zeta} ~+~ \frac{Z_m}{2 g \zeta(g) 
Z_\zeta} \sigma A^a_\mu A^{a \, \mu} ~-~ \frac{Z_m^2}{8\zeta(g) Z_\zeta} 
\left( A^a_\mu A^{a \, \mu} \right)^2 
\label{efflag}
\end{eqnarray} 
where 
\begin{equation}
e^{-W[J]} ~=~ \int {\cal D} A_\mu {\cal D} \psi {\cal D} \bar{\psi} {\cal D} c 
{\cal D} \bar{c} {\cal D} \sigma \, \exp \left[ \int d^d x \left( L^\sigma ~-~ 
\frac{\sigma J}{g} \right) \right] 
\end{equation}
which is now linear in the source term. Here the quantity $\zeta(g)$ denotes 
the explicit expression (\ref{zetaval}) and the extra terms again do not affect
the renormalization of (\ref{lag}) which retains the usual renormalization 
constants. The sigma field corresponds to the composite operator from its 
equation of motion and is similar to the $\sigma$ field of the Gross-Neveu 
model where it is introduced to remove the four fermion interaction. In 
(\ref{efflag}) a new renormalization constant, $Z_\zeta$, appears which is 
defined to be
\begin{equation}
Z_\zeta ~=~ 1 ~+~ \frac{\delta \zeta}{\zeta(g)}
\end{equation}
with $\zeta(g)$ also given by (\ref{zetaval}). To the order we will require
we note the value of the inverse, since that is what appears in (\ref{efflag}),
is 
\begin{eqnarray}
Z_\zeta^{-1} &=& 1 ~+~ \left( \frac{13}{6} C_A - \frac{4}{3} T_F \Nf \right) 
\frac{g^2}{16 \pi^2 \epsilon} \nonumber \\ 
&& +~ \left[ \left( 1464 C_A^2 T_F \Nf - 1365 C_A^3 - 384 C_A T_F^2 \Nf^2 
\right) \frac{1}{\epsilon^2} \right. \nonumber \\
&& \left. ~~~~~+~ \left( 5915 C_A^3 - 6032 C_A^2 T_F \Nf - 1248 C_A C_F T_F \Nf 
+ 1472 C_A T_F^2 \Nf^2 \right. \right. \nonumber \\
&& \left. \left. ~~~~~~~~~~~+~ 768 C_F T_F^2 \Nf^2 \right) \frac{1}{\epsilon} 
\right] \frac{g^4}{6144 \pi^4 (35 C_A-16 T_F \Nf)} ~+~ O(g^6) ~. 
\end{eqnarray}

We have now extended the LCO formalism of \cite{11} in the context of QCD and
$L^\sigma$, (\ref{efflag}), has emerged as the effective action to compute the 
effective potential as a function of $\sigma$. If $\sigma$ has a zero vacuum 
expectation value then one is in the usual perturbative QCD situation where the 
vacuum is known to be unstable, \cite{1,2,3,4}. However, if $\sigma$ has a 
non-zero vacuum expectation value it generates a gluon mass in a 
non-perturbative fashion in a vacuum whose stability properties need to be 
checked. This requires the explicit form of the effective potential. 

\sect{Effective potential.}
With the previous formalism we are now in a position to construct the effective
potential of the $\sigma$ field. It is derived by using standard techniques
since the source for the operator is now implemented linearly in the 
generating functional. Beyond the tree approximation the calculation breaks 
into two pieces. Briefly the one loop correction is determined by summing a set
of $n$-point diagrams with constant $\sigma$ external field. Hence, to one
loop the potential is
\begin{equation}
V(\sigma) ~=~ \frac{\sigma^2}{2g^2 \zeta(g) Z_\zeta} ~+~ \frac{(d-1)N_A}{2} 
\int \frac{d^d k}{(2\pi)^d} \, \ln \left( k^2 ~+~ \frac{\sigma}{g\zeta(g)}
\right)  
\label{barepot1}
\end{equation}
where the logarithm emerges from summing the appropriate diagrams based on
(\ref{efflag}). However, the momentum integral is divergent and the infinity
which cancels it arises from the counterterm available from $Z_\zeta$ in the
tree term as indicated in our notation. Hence, expanding to the finite part one 
is left with the one loop effective potential  
\begin{eqnarray}
V(\sigma) &=& 
\frac{9N_A}{2} \lambda_1 \sigma^{\prime \, 2} \nonumber \\
&& +~ \left[ \frac{3}{64} \ln \left( \frac{g \sigma^\prime}{\bar{\mu}^2} 
\right)
+ C_A \left(
-~ \frac{351}{8} C_F \lambda_1 \lambda_2
+ \frac{351}{16} C_F \lambda_1 \lambda_3
- \frac{249}{128} \lambda_2
+ \frac{27}{64} \lambda_3
\right)
\right. \nonumber \\
&& \left. ~~~~~ 
+~ C_A^2 \left(
-~ \frac{81}{16} \lambda_1 \lambda_2
+ \frac{81}{32} \lambda_1 \lambda_3
\right)
\right. \nonumber \\
&& \left. ~~~~~ 
+ \left(
-~ \frac{13}{128}
- \frac{207}{32} C_F \lambda_2
+ \frac{117}{32} C_F \lambda_3
\right)
\right] \frac{g^2 N_A \sigma^{\prime \, 2}}{\pi^2} ~+~ O(g^4)  
\label{effpot1} 
\end{eqnarray} 
where we have set 
\begin{equation}
\lambda_1 ~=~ [13 C_A-8 T_F \Nf]^{-1} ~~,~~
\lambda_2 ~=~ [35 C_A-16 T_F \Nf]^{-1} ~~,~~
\lambda_3 ~=~ [19 C_A-8 T_F \Nf]^{-1}
\end{equation}
and
\begin{equation}
\sigma ~=~ \frac{9 N_A}{(13 C_A - 8 T_F \Nf)} \sigma^\prime ~. 
\end{equation} 
As a check on the result we note that (\ref{effpot1}) agrees with \cite{11} in 
the $\Nf$ $\rightarrow$ $0$ limit. To go to the next order we must piece in not
only the higher loop Feynman diagrams but also the counterterms arising from 
the one loop integral of (\ref{barepot1}). The latter is achieved by replacing
$\sigma/(g\zeta(g))$ in the logarithm in (\ref{barepot1}) by 
$\sigma Z_m/(g\zeta(g) Z_\zeta)$. The resulting double and simple poles in 
$\epsilon$ will cancel the same poles in the two loop corrections. These are 
determined by repeating the resummation of the one loop calculation with 
$n$-point $\sigma$ Green's function but now with one radiative correction 
included on each topology. However, it is well known that this procedure is 
equivalently reproduced by computing two loop vacuum bubble graphs constructed 
from a new effective Lagrangian founded on (\ref{efflag}). This is given by 
shifting the $\sigma$ field to a new field, $\tilde{\sigma}$, which has a zero 
vacuum expectation value, and is defined by
\begin{equation}
\sigma ~=~ \langle \sigma \rangle ~+~ \tilde{\sigma}  
\end{equation} 
and then dropping terms linear in $\tilde{\sigma}$ and the overall additive
constant \cite{44}. One, in principle does this for all fields in 
(\ref{efflag}). However, the gluon, ghost and quark fields all carry either a 
Lorentz or colour index which means that their vacuum expectation values are 
each zero and thus the shifted fields are equivalent to the original ones. The 
net result for (\ref{efflag}) is that one has a massive gluon with a mass 
proportional to $\langle \sigma \rangle$ and the interactions of this new 
action are effectively the same as before. Therefore, the two loop effective 
potential, in the Landau gauge, is deduced from (\ref{efflag}) with a gluon 
whose propagator is 
\begin{equation}
-~ \frac{1}{(k^2-m^2)} \left[ \eta^{\mu\nu} ~-~ \frac{k^\mu k^\nu}{k^2} \right]
\end{equation}
where
\begin{equation}
m^2 ~=~ \frac{\sigma}{g \zeta(g)} ~. 
\end{equation} 
In our case there are five graphs to compute. Two purely gluonic and three
which involve either ghost, quark or $\tilde{\sigma}$ propagators. From
(\ref{efflag}) the propagator of the latter is momentum independent. 

The explicit calculation of each graph involves the evaluation of two loop
vacuum Feynman diagrams with both massive and massless propagators. However,
the exact expressions for such two loop topologies with three different masses
is already known and given in, for example, \cite{45}. We have written a 
{\sc Form} programme to handle the substitution and determination to the finite
part for each of the five diagrams. The result for the sum, after substituting 
for $\zeta(g)$ is 
\begin{eqnarray}
&& \frac{N_A \sigma^{\prime \, 2} g^4}{\pi^4} 
\left[ ~-~ \frac{9C_A}{4096\epsilon^2} 
+ \left( \frac{\Nf T_F}{192} - \frac{37C_A}{1536} 
+ \frac{9C_A}{2048} \ln \left( \frac{ g \sigma^\prime}{\bar{\mu}^2} \right) 
\right) \frac{1}{\epsilon} 
+ \left(  \frac{29}{2304} + \frac{\zeta(2)}{128} \right) \Nf T_F 
\right. \nonumber \\
&& \left. ~~~~~~~~~~~~~~+~ \left( \frac{891}{8192} s_2 - \frac{2023}{36864} 
- \frac{5\zeta(2)}{2048} \right) C_A 
- \frac{\Nf T_F}{96} \ln \left( \frac{ g \sigma^\prime}{\bar{\mu}^2} \right)  
\right. \nonumber \\
&& \left. ~~~~~~~~~~~~~~+~ \frac{37C_A}{768} \ln \left( 
\frac{ g \sigma^\prime}{\bar{\mu}^2} \right) 
- \frac{9C_A}{2048} \left( \ln \left( \frac{ g \sigma^\prime}{\bar{\mu}^2}
\right) \right)^2 \right] 
\label{twolp} 
\end{eqnarray}
where $s_2$ $=$ $(4 \sqrt{3}/3) Cl_2(\pi/3)$ and $Cl_2(x)$ is the Clausen 
function, $\zeta(n)$ is the Riemann zeta function and $\mu$ is the
renormalization scale introduced to ensure the coupling constant remains
dimensionless in $d$-dimensions. It is related to the usual $\MSbar$ scale 
$\bar{\mu}$ by $\bar{\mu}^2$ $=$ $4 \pi e^{-\gamma} \mu^2$ where $\gamma$ is 
the Euler-Mascheroni constant. Extending the one loop calculation to the finite
part of the next order in the coupling constant and adding to (\ref{twolp}) we 
find the finite two loop effective potential for non-zero $\Nf$ is 
\begin{eqnarray}
V(\sigma) &=& 
\frac{9N_A}{2} \lambda_1 \sigma^{\prime \, 2} \nonumber \\
&& +~ \left[ \frac{3}{64} \ln \left( \frac{g \sigma^\prime}{\bar{\mu}^2} 
\right)
+ C_A \left(
-~ \frac{351}{8} C_F \lambda_1 \lambda_2
+ \frac{351}{16} C_F \lambda_1 \lambda_3
- \frac{249}{128} \lambda_2
+ \frac{27}{64} \lambda_3
\right)
\right. \nonumber \\
&& \left. ~~~~~ 
+~ C_A^2 \left(
-~ \frac{81}{16} \lambda_1 \lambda_2
+ \frac{81}{32} \lambda_1 \lambda_3
\right)
+ \left(
-~ \frac{13}{128}
- \frac{207}{32} C_F \lambda_2
+ \frac{117}{32} C_F \lambda_3
\right)
\right] \frac{g^2 N_A \sigma^{\prime \, 2}}{\pi^2} 
\nonumber \\
&& +~ \left[ C_A 
\left(
-~ \frac{593}{16384}
- \frac{255}{16} C_F \lambda_2
+ \frac{36649}{4096} C_F \lambda_3
- \frac{1053}{64} C_F^2 \lambda_1 \lambda_2
+ \frac{1053}{128} C_F^2 \lambda_1 \lambda_3
\right. \right. \nonumber \\
&& \left. \left. ~~~~~~~~~~~
-~ \frac{5409}{1024} C_F^2 \lambda_2^2
+ \frac{1053}{1024} C_F^2 \lambda_3^2
+ \frac{891}{8192} s_2
- \frac{1}{4096} \zeta(2)
- \frac{3}{64} \zeta(3)
\right. \right. \nonumber \\
&& \left. \left. ~~~~~~~~~~~
+~ \frac{585}{16} \zeta(3) C_F \lambda_2
- \frac{4881}{256} \zeta(3) C_F \lambda_3
\right)
\right. \nonumber \\
&& \left. ~~~~~
+~ C_A^2 \left(
-~ \frac{11583}{128} C_F \lambda_1 \lambda_2
+ \frac{11583}{256} C_F \lambda_1 \lambda_3
+ \frac{72801}{2048} C_F \lambda_2^2
+ \frac{11583}{2048} C_F \lambda_3^2
\right. \right. \nonumber \\
&& \left. \left. ~~~~~~~~~~~~~~~
+~ \frac{3159}{128} C_F^2 \lambda_1 \lambda_2^2
+ \frac{3159}{512} C_F^2 \lambda_1 \lambda_3^2
+ \frac{372015}{16384} \lambda_2
- \frac{189295}{16384} \lambda_3
\right. \right. \nonumber \\
&& \left. \left. ~~~~~~~~~~~~~~~
+~ \frac{3159}{16} \zeta(3) C_F \lambda_1 \lambda_2
- \frac{3159}{32} \zeta(3) C_F \lambda_1 \lambda_3
- \frac{1053}{16} \zeta(3) C_F \lambda_2^2
\right. \right. \nonumber \\
&& \left. \left. ~~~~~~~~~~~~~~~
-~ \frac{3159}{256} \zeta(3) C_F \lambda_3^2
- \frac{6885}{256} \zeta(3) \lambda_2
+ \frac{116115}{8192} \zeta(3) \lambda_3
\right)
\right. \nonumber \\
&& \left. ~~~~~
+~ C_A^3 \left(
\frac{34749}{256} C_F \lambda_1 \lambda_2^2
+ \frac{34749}{1024} C_F \lambda_1 \lambda_3^2
+ \frac{64071}{512} \lambda_1 \lambda_2
- \frac{64071}{1024} \lambda_1 \lambda_3
\right. \right. \nonumber \\
&& \left. \left. ~~~~~~~~~~~~~~~
-~ \frac{694449}{16384} \lambda_2^2
- \frac{64071}{8192} \lambda_3^2
- \frac{9477}{32} \zeta(3) C_F \lambda_1 \lambda_2^2
- \frac{9477}{128} \zeta(3) C_F \lambda_1 \lambda_3^2
\right. \right. \nonumber \\
&& \left. \left. ~~~~~~~~~~~~~~~
- \frac{37179}{256} \zeta(3) \lambda_1 \lambda_2
+ \frac{37179}{512} \zeta(3) \lambda_1 \lambda_3
+ \frac{12393}{256} \zeta(3) \lambda_2^2
+ \frac{37179}{4096} \zeta(3) \lambda_3^2
\right)
\right. \nonumber \\
&& \left. ~~~~~
+~ C_A^4 \left(
-~ \frac{192213}{1024} \lambda_1 \lambda_2^2
- \frac{192213}{4096} \lambda_1 \lambda_3^2
+ \frac{111537}{512} \zeta(3) \lambda_1 \lambda_2^2
+ \frac{111537}{2048} \zeta(3) \lambda_1 \lambda_3^2
\right)
\right. \nonumber \\
&& \left. ~~~~~
+ \left(
-~ \frac{247}{4096} C_F
+ \frac{1185}{1024} C_F^2 \lambda_2
- \frac{615}{1024} C_F^2 \lambda_3
+ \frac{1}{128} \zeta(2) \Nf T_F
+ \frac{3}{64} \zeta(3) C_F
\right)
\right. \nonumber \\
&& \left. ~~~~~
+~ \left[ C_A \left(
+~ \frac{75}{4096}
- \frac{315}{1024} C_F \lambda_2
\right)
+ C_A^2 \left(
+ \frac{315}{4096} \lambda_2
\right)
+ \frac{9}{1024} C_F
\right] \ln \left( \frac{g \sigma^\prime}{\bar{\mu}^2} \right)
\right. \nonumber \\
&& \left. ~~~~~
-~ \frac{9}{4096} C_A \left( \ln \left( \frac{g \sigma^\prime}{\bar{\mu}^2} 
\right) \right)^2  
\right] \frac{g^4 N_A \sigma^{\prime \, 2}}{\pi^4} ~+~ O(g^6)  
\label{effpot2} 
\end{eqnarray}
which is the main result of this article. Again it agrees with the corrected
potential of \cite{11} when $\Nf$ $=$ $0$ which is a strong check on our
computations and renormalization though it is more complicated. 

With the explicit form we can make some general observations. One issue which 
always arises with potentials is that of boundedness. For a well defined 
physically interesting potential it must be bounded from below. However, the 
tree term is clearly dependent on the parameter $\lambda_1$ and therefore 
before any analysis can proceed we require  
\begin{equation} 
\lambda_1 ~=~ [13 C_A-8 T_F \Nf]^{-1} ~>~ 0 
\label{potcond}
\end{equation} 
as $N_A$ $>$ $0$. Interestingly this combination of group parameters is 
proportional to the one loop coefficient of the gluon wave function anomalous 
dimension in the Landau gauge. This would suggest that it has to be positive 
for a bounded potential and clearly this {\em cannot} occur in the case of 
quantum electrodynamics, (QED). Moreover, there has been renormalization group 
studies of the relation of this particular coefficient of the gluon anomalous 
dimension in connection with confinement, \cite{46,47}. Thus, it is curious 
that a positivity condition emerges on the one loop Landau gauge anomalous 
dimension in the effective potential of a composite operator which has the 
dimensions of mass since the presence of a mass gap in QCD is indicative of 
confinement. The other main point about (\ref{effpot2}) is that we have 
computed the two loop corrections with {\em massless} quarks in the one two 
loop diagram where quarks can appear. As quarks are massive in the real world 
including a mass in the quark propagator in fact leads to divergent terms 
proportional to the quark mass which cannot be cancelled off by terms in the 
current effective potential. One might expect that a mechanism similar to that 
of the LCO technique could be used to introduce mass terms for the quarks. 
However, that approach runs into immediate difficulties since a four quark 
interaction, analogous to the four gluon term of (\ref{efflag}), would emerge 
which would clearly lead to a non-renormalizable effective action. 
 
\sect{Analysis.}
With the explicit two loop effective potential for arbitrary colour group we
can now analyse various theories of interest. In particular we will focus on
the $SU(2)$ and $SU(3)$ colour groups. As the latter is strictly the theory of 
quarks in the nucleons we will provide this analysis in detail. First, to 
appreciate how our analysis is carried out we consider the one loop potential 
for an arbitrary gauge group. From (\ref{effpot1}) the first task is to 
determine the stationary points defined by the solution of 
\begin{equation}
\frac{d V}{d \sigma} ~=~ 0
\end{equation}
which implies  
\begin{eqnarray}
0 &=& N_A \sigma^\prime \left[  
\frac{9}{2} \lambda_1 ~+~ \left( \frac{3}{64} \ln \left( 
\frac{g \sigma^\prime}{\bar{\mu}^2} \right) + \frac{3}{128} 
+ \left(
-~ \frac{13}{128}
- \frac{207}{32} C_F \lambda_2
+ \frac{117}{32} C_F \lambda_3
\right)
\right. \right. \nonumber \\
&& \left. \left. ~~~~~~~~~~~~~~~~~~~~~~ 
+ C_A \left(
-~ \frac{351}{8} C_F \lambda_1 \lambda_2
+ \frac{351}{16} C_F \lambda_1 \lambda_3
- \frac{249}{128} \lambda_2
+ \frac{27}{64} \lambda_3
\right)
\right. \right. \nonumber \\
&& \left. \left. ~~~~~~~~~~~~~~~~~~~~~~ 
+~ C_A^2 \left(
-~ \frac{81}{16} \lambda_1 \lambda_2
+ \frac{81}{32} \lambda_1 \lambda_3
\right)
\right) \frac{g^2}{\pi^2} \right] ~+~ O(g^4) ~. 
\end{eqnarray}
Setting the scale to be that given where the logarithm vanishes, 
$g \sigma^\prime$ $=$ $\bar{\mu}^2$, then there are stationary points at 
\begin{equation}
\sigma^\prime ~=~ 0
\end{equation}
and 
\begin{equation}
\sigma^\prime ~\equiv~ \sigma^\prime_0 ~=~ \frac{\bar{\mu}^2}{g_0} 
\end{equation}
where 
\begin{equation} 
y_0 ~ \equiv ~ \frac{C_A g_0^2}{16\pi^2} ~=~ \frac{36 C_A 
( 35 C_A - 16 T_F \Nf )}{(6545 C_A^2 - 6008 C_A T_F \Nf + 864 C_F T_F \Nf 
+ 1280 T_F^2 \Nf^2 )} ~. 
\label{oneloopcc} 
\end{equation}  
This particular combination of the coupling constant on the left hand side has
been chosen in order to compare with \cite{11}. Clearly the trivial solution is 
the usual perturbative vacuum which has $V(0)$ $=$ $0$ whereas the latter 
induces a mass term for the gluon and is a global minimum provided 
(\ref{potcond}) is satisfied. With (\ref{oneloopcc}) we can obtain an $\Nf$ 
dependent estimate for the {\em effective} gluon mass, 
$m_{\mbox{\footnotesize{eff}}}$, defined in \cite{11} as  
\begin{equation} 
m_{\mbox{\footnotesize{eff}}}^2 ~=~ g \sigma^\prime ~.
\label{meffdef}
\end{equation} 
First, we need to convert our parameters to a reference mass scale, which is
taken to be $\Lambda_{\mbox{\footnotesize{$\MSbar$}}}$ and is introduced from
the solution of the one loop QCD $\beta$-function which determines the $\mu$
dependence of the running coupling constant, $g(\mu)$,  
\begin{equation} 
\frac{g^2(\mu)}{16\pi^2} ~=~ \left[2 \beta_0 \ln \left[ \frac{\mu^2}
{\Lambda^2_{\mbox{\footnotesize{$\MSbar$}}}}\right] \right]^{-1}  
\end{equation} 
where 
\begin{equation} 
\beta_0 ~=~ \frac{11}{3} C_A ~-~ \frac{4}{3} T_F \Nf ~.  
\end{equation} 
Hence, at one loop we obtain 
\begin{equation}
m_{\mbox{\footnotesize{eff}}} ~=~ 
\Lambda^{(\Nf)}_{\mbox{\footnotesize{$\MSbar$}}} \exp \left[ \frac{\left( 
6545 C_A^2 - 6008 C_A T_F \Nf + 864 C_F T_F \Nf + 1280 T_F^2 \Nf^2 \right)} 
{24( 11 C_A - 4 T_F \Nf )( 35 C_A - 16 T_F \Nf )} \right]
\end{equation}
which agrees with the one loop expression of \cite{11} for the unitary colour 
groups when $\Nf$ $=$ $0$. It is worth noting that this formula is {\em not} 
valid in the QED case where $C_A$~$=$~$0$, $C_F$~$=$~$1$ and $T_F$~$=$~$1$ 
because the original one loop potential is then unbounded from below.  

Repeating this analysis for the potential at two loops for an arbitrary colour
group is extremely cumbersome and it is more appropriate to concentrate on 
specific interesting cases. For instance, for $SU(3)$ the two loop potential 
becomes 
\begin{eqnarray}
\left. V(\sigma) \right|_{SU(3)} &=& \frac{36}{(39 - 4 \Nf)}
\sigma^{\prime \, 2} \nonumber \\
&& +~ \left[ \frac{3}{8} \ln \left( \frac{g \sigma^{\prime \, 2}}{\bar{\mu}^2} 
\right) 
- \frac{13}{16}
+ \frac{3537}{4 (39 - 4 \Nf) (57 - 4 \Nf)}
- \frac{3537}{2 (39 - 4 \Nf) (105 - 8 \Nf)}
\right. \nonumber \\
&& \left. ~~~~~
+~ \frac{393}{8 (57 - 4 \Nf)}
- \frac{1851}{16 (105 - 8 \Nf)}
\right] \frac{g^2 \sigma^{\prime \, 2}}{\pi^2} 
\nonumber \\
&& 
+~ \left[
\frac{2673}{1024} s_2
- \frac{9289}{6144}
- \frac{3}{512} \zeta(2)
+ \frac{1}{32} \zeta(2) \Nf
- \frac{5}{8} \zeta(3)
\right. \nonumber \\
&& \left. ~~~~~
-~ \frac{1129005}{128 (39 - 4 \Nf) (57 - 4 \Nf)}
+ \frac{397305}{64 (39 - 4 \Nf) (57 - 4 \Nf)} \zeta(3)
\right. \nonumber \\
&& \left. ~~~~~
-~ \frac{10161045}{512 (39 - 4 \Nf) (57 - 4 \Nf)^2}
+ \frac{3575745}{256 (39 - 4 \Nf) (57 - 4 \Nf)^2} \zeta(3)
\right. \nonumber \\
&& \left. ~~~~~
+~ \frac{1129005}{64 (39 - 4 \Nf) (105 - 8 \Nf)}
- \frac{397305}{32 (39 - 4 \Nf) (105 - 8 \Nf)} \zeta(3)
\right. \nonumber \\
&& \left. ~~~~~
-~ \frac{10161045}{128 (39 - 4 \Nf) (105 - 8 \Nf)^2}
+ \frac{3575745}{64 (39 - 4 \Nf) (105 - 8 \Nf)^2} \zeta(3)
\right. \nonumber \\
&& \left. ~~~~~
-~ \frac{3404293}{6144 (57 - 4 \Nf)}
+ \frac{420267}{1024 (57 - 4 \Nf)} \zeta(3)
- \frac{1129005}{1024 (57 - 4 \Nf)^2}
\right. \nonumber \\
&& \left. ~~~~~
+~ \frac{397305}{512 (57 - 4 \Nf)^2} \zeta(3)
+ \frac{7012085}{6144 (105 - 8 \Nf)}
- \frac{24525}{32 (105 - 8 \Nf)} \zeta(3)
\right. \nonumber \\
&& \left. ~~~~~
-~ \frac{12222795}{2048 (105 - 8 \Nf)^2}
+ \frac{132435}{32 (105 - 8 \Nf)^2} \zeta(3)
\right. \nonumber \\
&& \left. ~~~~~
+ \left( \frac{273}{512}
- \frac{2205}{512 (105 - 8 \Nf)}
\right)
\ln \left( \frac{g \sigma^{\prime \, 2}}{\bar{\mu}^2} \right) 
\right. \nonumber \\
&& \left. ~~~~~
-~ \frac{27}{512}
\left( \ln \left( \frac{g \sigma^{\prime \, 2}}{\bar{\mu}^2} \right) \right)^2 
\right] \frac{g^4 \sigma^{\prime \, 2}}{\pi^4} ~+~ O(g^6) ~. 
\label{effpot23}
\end{eqnarray}
{\begin{table}[ht] 
\begin{center} 
\begin{tabular}{|c||c|c|} 
\hline
$N_{\!f}$ & $y^{(1)}_0$ & $y^{(2)}_0$ \\ 
\hline
 0 & 0.1925134 & 0.1394790 \\ 
 2 & 0.2219195 & 0.1832721 \\ 
 3 & 0.2398224 & 0.2225696 \\ 
\hline
\end{tabular} 
\end{center} 
\begin{center} 
{Table 1. One and two loop estimates of the coupling constant at the minimum
of $V(\sigma)$ for $SU(3)$.}
\end{center} 
\end{table}}  
With (\ref{effpot23}) we can determine the value of the coupling constant where
the minimum occurs. These are recorded in Table 1 where we concentrate on the 
values $\Nf$ $=$ $2$ and $3$ and the superscript here and elsewhere denotes the
loop order. To consider more flavours one would be in the situation where the 
charm quark is present but for the case where it is treated as massless. As its
mass is more comparable with the one loop effective gluon mass than the three 
light quarks and since we have neglected quark mass effects it is not clear how
it would influence the approximations we have made. 

We can now determine the effect the two loop corrections have on the effective 
gluon mass, $m_{\mbox{\footnotesize{eff}}}$. This requires the inclusion of the 
two loop correction to the solution of the $\beta$-function for the coupling 
constant as a function of $\mu$ which gives
\begin{equation} 
\frac{g^2(\mu)}{16\pi^2} ~=~ \left[2 \beta_0 \ln \left[ \frac{\mu^2}
{\Lambda^2_{\mbox{\footnotesize{$\MSbar$}}}}\right] \right]^{-1} \left[ 1 ~-~ 
\beta_1 \left[2\beta_0^2 \ln \left[ \frac{\mu^2}
{\Lambda^2_{\mbox{\footnotesize{$\MSbar$}}}}\right] \right]^{-1}  
\ln \left[ 2 \ln \left[ \frac{\mu^2}
{\Lambda^2_{\mbox{\footnotesize{$\MSbar$}}}}\right] \right] \right] 
\end{equation} 
where
\begin{equation} 
\beta_1 ~=~ \frac{34}{3} C_A^2 ~-~ 4 C_F T_F \Nf ~-~ \frac{20}{3} 
C_A T_F \Nf ~.
\end{equation} 
{\begin{table}[hb] 
\begin{center} 
\begin{tabular}{|c||c|c|} 
\hline
$N_{\!f}$ & $m^{(1)}_{\mbox{\footnotesize{eff}}}/ 
\Lambda^{(\Nf)}_{\mbox{\footnotesize{$\MSbar$}}}$ & 
$m^{(2)}_{\mbox{\footnotesize{eff}}}/
\Lambda^{(\Nf)}_{\mbox{\footnotesize{$\MSbar$}}}$ \\ 
\hline
 0 & 2.03 & 2.12 \\ 
 2 & 2.01 & 1.99 \\ 
 3 & 2.00 & 1.89 \\ 
\hline
\end{tabular} 
\end{center} 
\begin{center} 
{Table 2. One and two loop estimates of the gluon effective mass for $SU(3)$.}
\end{center} 
\end{table}}  
Converting the coupling constant at the minimum of the potential, $g_0$, when
$g \sigma^\prime$ $=$ $\bar{\mu}^2$ and substituting into (\ref{meffdef}) we
obtain the one and two loop estimates given in Table 2. We have expressed our 
estimates in terms of the $\Nf$-dependent 
$\Lambda^{(\Nf)}_{\mbox{\footnotesize{$\MSbar$}}}$ which for low $\Nf$ can be 
taken to be $237$ MeV giving a two loop estimate of around $450$ MeV for $\Nf$ 
$=$ $3$. From the table, aside from the Yang-Mills result of \cite{11} at two 
loops it is clear this effective gluon mass decreases slowly with the increase 
in the number of quark flavours. However, it is not clear whether this a real 
feature since quark mass effects, which should not be significant for light 
flavours, have been neglected. Moreover, it is worth stressing that 
$m_{\mbox{\footnotesize{eff}}}$ is {\em not} the physical gluon mass in the new
vacuum. The physical mass can only be deduced by computing the pole of the 
gluon propagator at {\em two} loops using the effective action (\ref{efflag}).
This is clearly beyond the scope of the present article. 
{\begin{table}[ht]
\begin{center} 
\begin{tabular}{|c||c|c|} 
\hline
$N_{\!f}$ & $y^{(1)}_0$ & $y^{(2)}_0$ \\ 
\hline
 0 & 0.1925134 & 0.1394790 \\ 
 2 & 0.2416107 & 0.2239704 \\ 
 3 & 0.2758161 & 0.4342486 \\ 
\hline
\end{tabular} 
\end{center} 
\begin{center} 
{Table 3. One and two loop estimates of the coupling constant at the minimum
of $V(\sigma)$ for $SU(2)$.}
\end{center} 
\end{table}}  

We have repeated the above analysis for the case of $SU(2)$ which follows the
same path and the results are summarized in Tables $3$ and $4$. For 
$\Nf$~$=$~$3$ a second positive root emerged for the value for which the 
potential was a minimum at two loops. This was $y^{(2)}_0$~$=$~$0.7559862$ 
which we have neglected in the table as its value is significantly different 
from the one loop case. Though since the value given in Table 3 seems to be out
of step with the pattern of Table 1 it is not clear how reliable that coupling
is. The effective gluon mass values given in Table $4$ are derived using the 
couplings of Table $3$. Overall for most of the $SU(2)$ and $SU(3)$ cases we 
have analysed the two loop corrections to the effective gluon mass are no more 
than about $5\%$ of the one loop result which is encouraging since it indicates
a degree of stability of the approximation. However, the exception is the 
$\Nf$~$=$~$3$ $SU(2)$ case which gives a gluon mass estimate which is $20\%$ 
different from the one loop estimate. As in this case a second positive root 
emerged for the coupling constant at the minimum, it may require a three loop 
effective potential to establish which if either is spurious. 
{\begin{table}[ht] 
\begin{center} 
\begin{tabular}{|c||c|c|} 
\hline
$N_{\!f}$ & $m^{(1)}_{\mbox{\footnotesize{eff}}}/
\Lambda^{(\Nf)}_{\mbox{\footnotesize{$\MSbar$}}}$ & 
$m^{(2)}_{\mbox{\footnotesize{eff}}}/
\Lambda^{(\Nf)}_{\mbox{\footnotesize{$\MSbar$}}}$ \\ 
\hline
 0 & 2.03 & 2.12 \\ 
 2 & 1.99 & 1.88 \\ 
 3 & 1.97 & 1.58 \\ 
\hline
\end{tabular} 
\end{center} 
\begin{center} 
{Table 4. One and two loop estimates of the gluon effective mass for $SU(2)$.}
\end{center} 
\end{table}}  

Finally, following the method of \cite{31}, we have computed the value of the 
one and two loop effective potentials at the non-trivial minimum and presented 
the numerical values in Table $5$ for $SU(3)$ where they are all clearly 
negative. As the potentials are zero at the trivial minimum this justifies our 
remark that the non-trivial minimum corresponds to the stable vacuum. Clearly 
the inclusion of quarks increases the value of the minimum but the two loop 
correction is significantly larger. 
{\begin{table}[hb] 
\begin{center} 
\begin{tabular}{|c||c|c|} 
\hline
$N_{\!f}$ & $V^{(1)}_{\mbox{\footnotesize{min}}} 
/\left( \Lambda^{(\Nf)}_{\mbox{\footnotesize{$\MSbar$}}} \right)^4$ & 
$V^{(2)}_{\mbox{\footnotesize{min}}} 
/\left( \Lambda^{(\Nf)}_{\mbox{\footnotesize{$\MSbar$}}} \right)^4$ \\ 
\hline
 0 & $-$ $0.32300$ & $-$ $0.76451$ \\ 
 2 & $-$ $0.31145$ & $-$ $0.66966$ \\ 
 3 & $-$ $0.30617$ & $-$ $0.61079$ \\ 
\hline
\end{tabular} 
\end{center} 
\begin{center} 
{Table 5. One and two loop estimates of the minimum of the effective potential
for $SU(3)$.}
\end{center} 
\end{table}}  

\sect{Discussion.}
We conclude with various remarks. First, we have included quarks in the 
effective potential of the dimension two operator, $\half A^{a \, 2}_\mu$ in 
the Landau gauge and shown that the potential is bounded from below for QCD but
not for QED. Also, we have demonstrated that an effective gluon mass decreases 
in value slowly as $\Nf$ increases. However, this is for the unrealistic case 
of massless quarks, though for the values of $\Nf$ we have studied the quarks 
are light and their mass effects ought not to be significant. Clearly, for a 
more 
pragmatic study it would be better if quark masses could be included but this 
would require the removal of extra infinities which are quark mass dependent 
from the {\em two} loop effective potential. At present this will require an 
extension of the LCO formalism or another mechanism. Moreover, whilst our gluon
effective mass estimates are essentially for a mass which from $L^\sigma$ 
corresponds to a classical mass, a much better quantity to study would be the 
physical gluon mass itself which is defined as the pole of the gluon propagator
derived from the effective action by computing the radiative corrections to the
gluon two-point function. In the application of the LCO technique to the 
Gross-Neveu model and after improving convergence of the series, \cite{32,33},
it was shown that such a physical mass was accurate to a few percent with the 
known exact mass gap. It would be hoped that a similar calculation for QCD 
would give a reliable estimate for the gluon mass. Finally, given that we have 
now derived a two loop effective potential for a particular dimension two 
composite operator there is scope for examining the potentials of other related
dimension two operators. For instance, ghost number breaking operators have 
been considered by various authors either in the maximal abelian gauge or in 
the Curci-Ferrari gauge which acts as a testbed for understanding gluon mass 
generating problems in the former gauge. Since we have simplified the way of 
computing the quantity $\zeta(g)$ without having to resort to the computation 
of three loop massive vacuum bubble graphs then the construction of such 
potentials for other operators or combination of operators can proceed in an 
efficient manner. 

\vspace{1cm} 
\noindent
{\bf Acknowledgement.} This work was supported in part by {\sc PPARC} through
a research studentship, (REB). We also thank H. Verschelde and D. Dudal for 
discussions.  


\end{document}